\documentclass{PoS}
\usepackage{multirow}
\usepackage{amsmath}
\usepackage{graphicx}

\usepackage{slashed}

\newcommand{\tr}{\text{tr}\,}

\title{Effective Polyakov loop models for QCD-like theories at finite chemical potential}

\ShortTitle{Effective Polyakov loop models for QCD-like theories at finite chemical potential}

\author{\speaker{Philipp Scior}\\
		Theoriezentrum, Institut f\"ur Kernphysik, TU Darmstadt, 64289 Darmstadt, Germany\\
        E-mail: \email{scior@theorie.ikp.physik.tu-darmstadt.de}}

\author{Lorenz von Smekal\\
         Institut f\"ur Theoretische Physik, Justus-Liebig-Universit\"at Gie{\ss}en, 35392 Gie{\ss}en, Germany\\
       E-mail: \email{lorenz.v.Smekal@theo.physik.uni-giessen.de}}

\abstract{We study genuine finite density effects in QCD-like theories with three-dimensional Polyakov-loop effective theories for heavy quarks. These are derived from the full QCD-like theories by combined strong-coupling and hopping expansions. In particular, we investigate the cold and dense regimes of phase diagrams where we expect to find Bose-Einstein-condensation of diquark baryons or a fermionic first-order liquid-gas transition, depending on the gauge group of the theory. In two-color QCD, for example, we observe evidence of a continuous zero-temperature transition to finite diquark density when the quark chemical potential $\mu$ reaches half the diquark mass, i.e. without binding energy. In G$_2$-QCD we observe, in addition to this ``Silver Blaze'' onset of diquark density, a second transition in the density towards an exponential increase by roughly $3\mu/T$ corresponding to a finite density of G$_2$-nucleons.}

\FullConference{The 33rd International Symposium on Lattice Field Theory\\
		14 -18 July 2015\\
		Kobe International Conference Center, Kobe, Japan}

\begin{document}

\section{Introduction}
Much progress towards the exploration of the QCD phase diagram was made in recent years. However, the low temperature and high baryon-density region of the phase diagram remains challenging for first principles studies because the fermion sign problem impedes standard lattice Monte-Carlo simulations \cite{deForcrand:2010ys}. One way around this problem is to use effective theories in terms of Polyakov-loop variables which can be derived by combined strong-coupling and hopping expansions \cite{Fromm:2012eb,Langelage:2014vpa}. Here, we give an update on our project to apply such expansions to QCD-like theories without sign problem, e.g. two-color QCD or G$_2$-QCD, where the results form effective theories can thus be confronted with lattice simulations of the full theory at finite baryon density.

The phase diagram of two-color QCD has long been studied in effective chiral models \cite{Kogut:2000ek} and with lattice simulations \cite{Cotter:2012mb}. A second-order phase transition to a Bose-Einstein condensate occurs at zero temperature when the baryon chemical potential $\mu_B $ reaches the mass of the bosonic diquark $m_d$.
G$_2$-QCD contains bosonic as well as fermionic baryons. Here lattice simulations have shown evidence of both, a transition to finite diquark density as in two-color QCD followed by a first-order transition to what one might call the analogue of nuclear matter in G$_2$-QCD \cite{Wellegehausen:2013cya}. 

\section{Effective Polyakov Loop Theory}
We will summarize the derivation of the effective action only very briefly. We rather refer to a very detailed description for SU(3) in \cite{Langelage:2014vpa} which is similar to the case of other gauge groups. Here we will sketch the derivation of the effective action for two-color QCD with $N_f$ flavors of Wilson quarks which is defined by integrating out the spatial links in the partition function 
\begin{equation}
Z =\int [dU_0] \; \exp[-S_\text{eff}] \; , \qquad 
\exp[-S_\text{eff}]= \int [dU_ i] \; \exp[-S_\text{g}] \prod_{f=1}^{N_f} \det D_f \; . \label{eff_action}
\end{equation}
It can the be separated in two parts, $	S_\text{eff}(W)= S_1(W) + S_2(W)$, which both depend on temporal Wilson lines $W$  or their traces, the Polyakov loops,
\begin{equation}
L_{\vec x}= \tr W_{\vec x} =\tr \prod_{t=0}^{N_t-1} U_0(\vec x,t) \; .
\end{equation}
$S_1$ comes from the gauge action of the theory, with effective couplings modified by fermionic contributions. $S_2$ contains fermionic loops that wind around the time direction of the lattice. 

The most convenient way to derive $S_1$ is by character expansion of $\exp[-S_g]$. However, at the low temperatures we are interested in here, on lattices with a very large number of time slices $N_t > 100 $ the effective couplings $\lambda$ in $S_1$ are all very small. For two-color QCD the leading one results to be $\lambda(\beta=2.5,N_t=200)\sim 10^{-15}$ at the order $\beta^{12}$ for $\beta = 2.5$ and temperatures around $T\sim 12 $ MeV, for example. For even lower $T$ it continues to decrease exponentially and we therefore neglect $S_1$ altogether \cite{Scior:2015vra}. 
$S_2$ can be obtained from a hopping expansion of the fermion determinant which one factorizes in a static part, containing only contributions from temporal hops, and a kinetic part with all the remaining ones involving spatial hops,
\begin{align}
\det[D]&=\det[D_\text{stat}]\det[D_\text{kin}] \; . \label{determinant}
\end{align}
In the strong-coupling limit $\beta=0$ the static determinant is readily evaluated because no integration over spatial links is required, it factorizes in flavor, spin and space,
\begin{equation}
 	\det[D_\text{stat}]=\prod_{\vec x} (1+hL_{\vec x} +h^2)^{2 N_f}(1+\bar h L_{\vec x}  + \bar h^2)^{2 N_f} \; ,
 \end{equation} 
 with the effective fermion couplings
 \begin{equation}
 	h=(2 \kappa e^{a \mu})^{N_t}=e^{\frac{\mu-m_q}{T}} \hspace{0.2cm} \text{and} \hspace{0.2cm} \bar h = h(-\mu) \; . \label{ferm_coupling}
 \end{equation}
Gauge corrections in terms of the fundamental expansion parameter  $u= I_2(\beta)/I_1(\beta)$ of the character expansion of $\exp[-S_g]$ lead to modifications of the effective coupling $h$ of the form,
\begin{equation}
	h = \exp \left[ N_t \left(a \mu + \ln 2 \kappa + 6 \kappa^2 \frac{u}{1-u} + \, \cdots\right) \right] \; . \label{h1}
\end{equation}
These corrections correspond to a shift in the constituent quark mass
\begin{align}
  a m_q =& - \ln(2\kappa) - 6 \kappa^2 \frac{u}{1-u} + 48 \kappa^4 u  (1-u) - 24 \kappa^2 u \, (u^4  + \kappa^2 u^2 -  \kappa^4 ) + \, \cdots \; . \label{quarkmass}
\end{align}
The evaluation of the kinetic part of the fermion determinant is a little more complicated. One obtains terms with 2-point and 3-point interactions between temporal Wilson lines. With these the effective action up to the order $\kappa^4$, including SU(2) specific diquark contributions at this order,
altogether reads \cite{Scior:2015vra}
\small
\begin{align}
  -S_\text{eff}=&N_f \sum_{\vec x} \log (1+hL_{\vec x} +h^2)^2-2N_f h_2 \sum_{\vec x, i} \tr \frac{h W_{\vec x}}{1 +h W_{\vec x}} \tr \frac{h W_{\vec x+i}}{1+h W_{\vec x+i}} \notag \\
  &  \hskip -1.2cm
  +2N_f^2 \frac{\kappa^4 N_t^2}{N_c^2} \sum_{\vec x, i} \tr \frac{h W_{\vec x}}{(1 +h W_{\vec x})^2} \tr \frac{h W_{\vec x+i}}{(1+h W_{\vec x+i})^2}
  +N_f \frac{\kappa^4 N_t^2}{N_c^2} \sum_{\vec x, i, j} \tr \frac{h W_{\vec x}}{(1 +h W_{\vec x})^2} \tr \frac{h W_{\vec x-i}}{1+h W_{\vec x-i}} \tr \frac{h W_{\vec x-j}}{1+h W_{\vec x-j}} \notag \\
  & 
 \hskip -1.2cm
  +2N_f \frac{\kappa^4 N_t^2}{N_c^2} \sum_{\vec x, i, j} \tr \frac{h W_{\vec x}}{(1 +h W_{\vec x})^2} \tr \frac{h W_{\vec x-i}}{1+h W_{\vec x-i}} \tr \frac{h W_{\vec x+j}}{1+h W_{\vec x+j}}
  + N_f\frac{\kappa^4 N_t^2}{N_c^2} \sum_{\vec x, i, j} \tr \frac{h W_{\vec x}}{(1 +h W_{\vec x})^2} \tr \frac{h W_{\vec x+i}}{1+h W_{\vec x+i}} \tr \frac{h W_{\vec x+j}}{1+h W_{\vec x+j}} \notag \\
&+(2N_f^2-N_f)\kappa^4 N_t^2 \sum_{\vec x,i} \frac{h^4}{(1+h L_{\vec x}+h^2)(1+h L_{\vec x+i}+h^2)} \; , \label{eff_action_kappa4_Nf2}
\end{align}
\normalsize
where the second fermion coupling $h_2$ is defined as
\begin{equation}
	h_2= \frac{\kappa^2 N_t}{N_c}\left[1+2 \frac{u}{1-u}+ \, \cdots \right] \; . \label{h2}
\end{equation}
This is an already simplified version of the effective action valid for the cold and dense regime of the phase diagram where terms proportional to $\bar h $
and terms subleading in $N_t$ can be dropped.

\begin{figure}[t]
  \vskip -.2cm
\begin{minipage}{0.49\textwidth}
	\hskip -.2cm\includegraphics[width=1.1\linewidth]{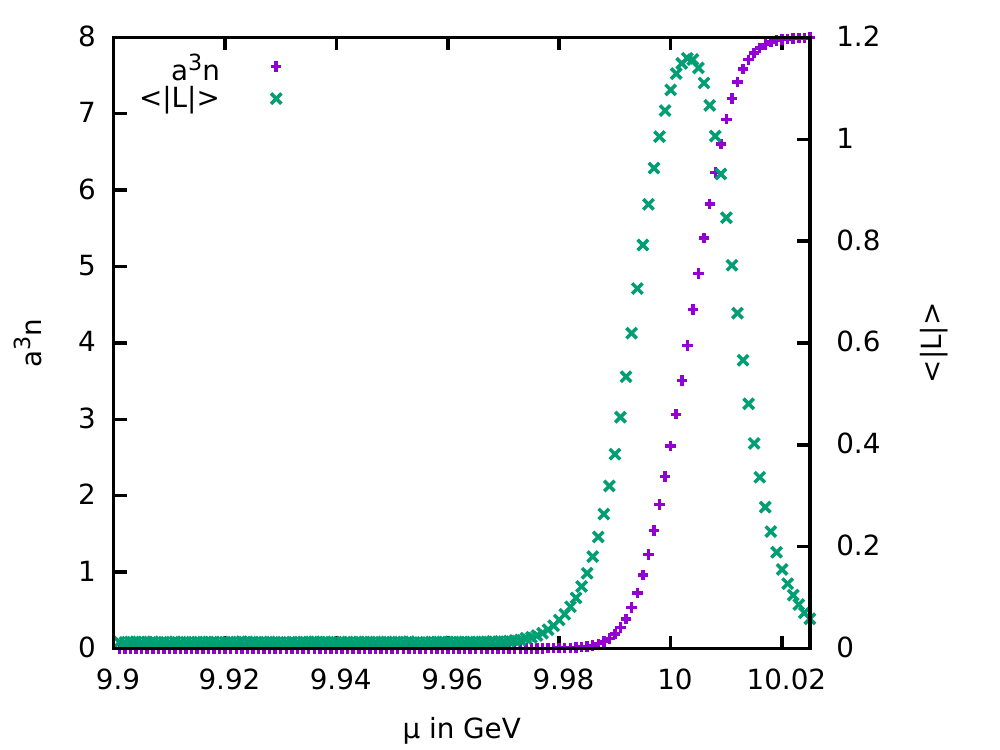}
	\caption{Quark density  $a^3n$ in lattice units and Polyakov-loop expectation value $\langle |L| \rangle$, both over $\mu$, with simulation parameters $\beta=2.5$, $\kappa=0.00802$, leading to $m_d=20$~GeV, $N_s=16$, $N_t=484$, corresponding to $T=5$ MeV, and $N_f=2$.\label{linear_dens_poly_Nf2} }

\end{minipage} \hspace*{0.1cm}
\begin{minipage}{0.49\textwidth}
	\vspace*{-0.5cm}
	\hskip -.2cm
	\includegraphics[width=1.1\linewidth]{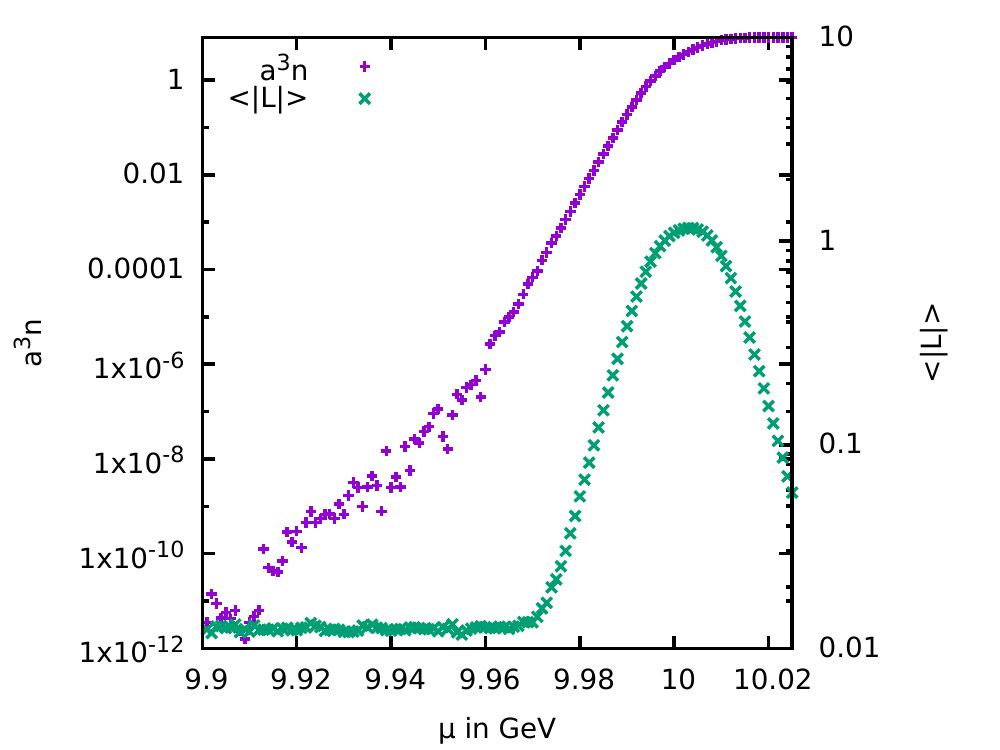}
	\caption{\label{log_dens_poly_Nf2} Logarithmic plot of the density in lattice units $a^3n$ and the Polyakov loop $\langle |L| \rangle$ as a function of $\mu$ in physical units  at $T=5$ MeV with the same parameters as in Fig.~1.}
\end{minipage}
\end{figure}

\section{Results}
To set the scale we use the non-perturbative $\beta$-function for the pure SU(2) gauge theory from \cite{Smith:2013msa} with the typical $\sqrt{\sigma}=440$ MeV for the string tension. The hopping parameter $\kappa $ is then used to adjust  the diquark mass via the combined strong-coupling/hopping expansion formula,
\begin{equation}
 a m_d= - 2 \ln(2\kappa) - 6 \kappa^2 -24 \kappa^2 \frac{u}{1-u} + 6 \kappa^4 +\, \cdots \; , \label{mass}
\end{equation}
Note that the relevant expansion parameter in the effective action is $\kappa^2 N_t/N_c$. Because we are interested in very low temperatures $T = 1/a N_t$ we need large $N_t$. Therefore, our hopping parameter $\kappa$ needs to be sufficiently small and our quarks and diquarks sufficiently heavy. In all our simulations we have fixed the diquark mass to $m_d=20$ GeV with temperatures ranging between between $T= 3.454$ MeV and $9$ MeV. On our finest lattice with $\beta = 2.5$, corresponding to $a=0.0810 $~fm, this amounts to $\kappa = 0.00802123$ and $N_t$ values between $ 269$ and $700$. We have veryfied explicitly that this is within the range of validity of the expansion in the effective action \cite{Scior:2015vra}.

In Fig.~\ref{linear_dens_poly_Nf2} we show the quark density and the Polyakov loop in the effective theory with $N_{f}=2$ degenerate quark flavors from this  $\beta = 2.5$ lattice, with $N_s = 16$ and $N_t=484$, corresponding to $T= 5$~MeV. Because of finite volume and quark mass, even for vanishing net-baryon density the Polyakov loop has a small but non-zero expectation value $\langle |L| \rangle > 0$. However, it remains constant at its $\mu =0$ value until just below the onset of baryon density expected at $m_d/2$. In the same region of quark chemical potentials around $\mu=m_d/2$ the quark-number density in lattice units $a^3n$ shows its most rapid increase before it saturates at $2 N_c N_f=8$, corresponding to the maximum number of quarks per site allowed by the Pauli principle. This artificial behavior on the lattice leads to an effective quenching of the quarks and hence the Polyakov loop decreases again as it is approached.

\begin{figure}[t]
\vspace*{-0.4cm}
\begin{minipage}{0.49\textwidth}
	\hskip -.1cm
	\includegraphics[width=\linewidth]{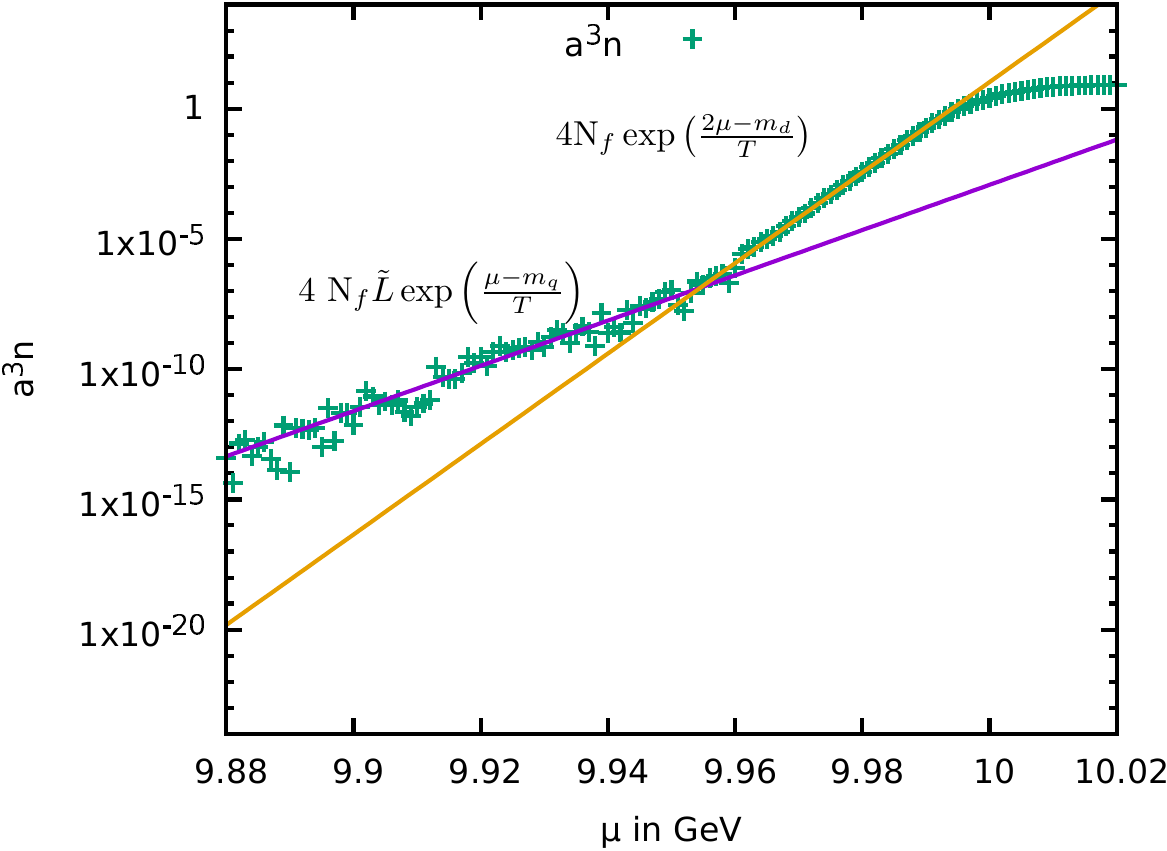}
	\caption{Logarithmic plot of the density in lattice units $a^3n$ at $T=5$ MeV compared to one-parameter fits using $\tilde L$ for $\mu < 9.96$~GeV, and $m_d$ for $\mu > 9.96$~GeV, see text. \label{slopes}}

\end{minipage} \hspace*{0.1cm}
\begin{minipage}{0.49\textwidth}
	\vspace*{-0.5cm}
	\hskip -.2cm
	\includegraphics[width=1.03\linewidth]{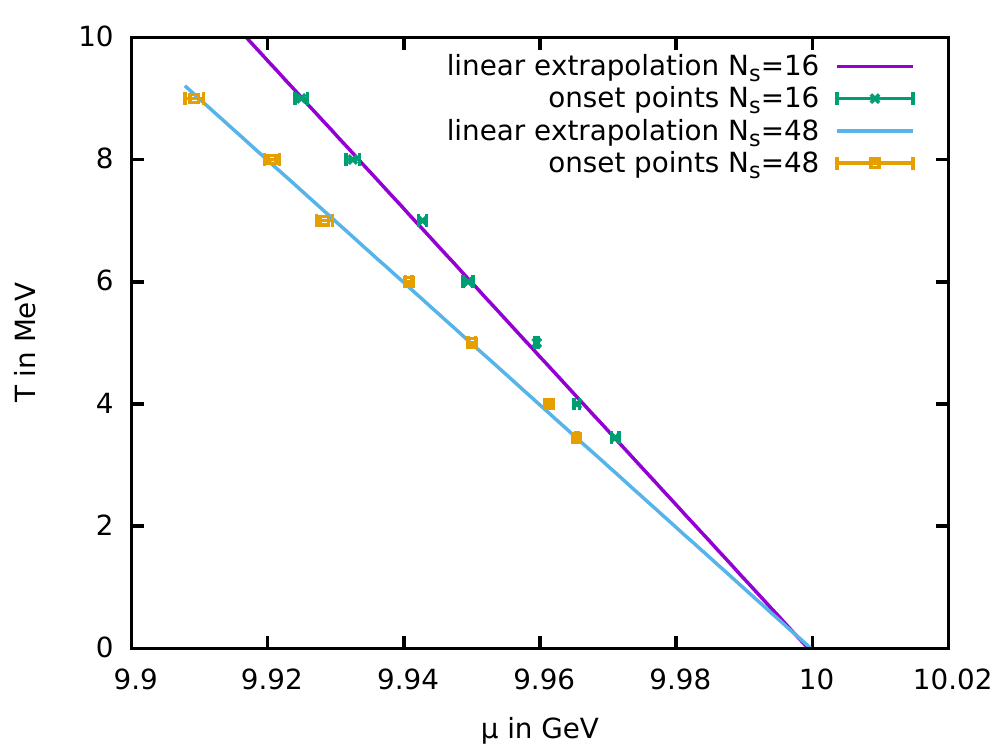}
	\caption{Section of the $N_f=2$ phase diagram with linear extrapolations of the diquark-density onset to $T=0$ at $\beta = 2.5$  with $N_s=16$ and 48. \label{extrapol}}
\end{minipage}
\end{figure}

In order to distinguish baryon density from quark density,  Fig.~\ref{log_dens_poly_Nf2} we show a logarithmic plot of the same data  as in Fig.~\ref{linear_dens_poly_Nf2}. This reveals two different regimes of exponential increase in the quark-number density before it eventually approaches lattice saturation.
They are separated by a kink at $\mu_c \approx 9.96$~GeV in the logarithmic plot (not visible on the scales of Fig.~\ref{linear_dens_poly_Nf2}), where the Polyakov-loop is still at its constant $\mu=0$ expectation value. The location of the kink and the overall behavior of the density are very well described by  a mean-field formula for the static fermion determinant
\begin{equation}
\frac{a^3 n}{4 N_f} =  \frac{1+\tilde Le^\frac{m_q-	\mu}{T}}{1+2\tilde Le^\frac{m_q-	\mu}{T}+e^\frac{2(m_q-	\mu)}{T}}\to \left\{\begin{array}{ll}  e^{(\mu_B - m_d)/T}\; , & \tilde L \, e^\frac{m_q-	\mu}{T} \ll 1 \; ,  \\[4pt]
 \tilde L\,  e^{(\mu - m_q)/T}\; , & \tilde L\,  e^\frac{m_q-	\mu}{T} \gg 1 \; ,
  \end{array} \right.   \label{mean-field_dens}
\end{equation}
where  $\tilde L \in [-1,1]$ represents a normalized Polyakov-loop expectation value. With the quark mass from Eq.~(\ref{quarkmass}) a one-parameter fit via  $\tilde L$ to the first exponential for $\mu < 9.96 $~GeV as in Fig.~\ref{slopes} yields a value consistent with the expectation value from the per-site probability distribution obtained by histogramming the local Polyakov-loop as in \cite{Smith:2013msa}.
When we fit the data in the region of the second exponential, for 9.96~GeV
$< \mu  < $ 9.99~GeV, using $m_\mathrm{fit}$  as a single fit parameter
in the place of $m_d$,
we obtain $m_\mathrm{fit} = 19.9986(10)$~GeV very close to the scalar diquark mass fixed at  $m_d = 20$~GeV, and significantly smaller than $2m_q=20.0028(2)$~GeV. We take the intersection of the two at $\mu_c = m_\mathrm{fit} - m_q + T \ln \tilde L$  as the transition point from a thermal gas of heavy quarks with imperfect statistical confinement in a finite volume to an essentially free gas of heavy diquarks.

Extracting this transition point for various temperatures from $9$ MeV down to $3.454$ MeV we can extrapolate the corresponding chemical potentials $\mu_c$ to $T=0$. Fig.~\ref{extrapol} shows the results of linear extrapolations for $N_s=16$ and 48 lattices. The smaller slope on the larger lattice reflects the volume dependence of the Polyakov loop. The $T=0$ diquark onsets are obtained as $\mu_c = 9.9994(18)$~GeV and 9.9998(9)~GeV, respectively, and hence include $m_d/2 = 10$~GeV within the errors.

	\begin{figure}[t]
\begin{minipage}{0.49\textwidth}
	\vspace*{-0.9cm}
	\hskip -.1cm
	\includegraphics[width=\linewidth]{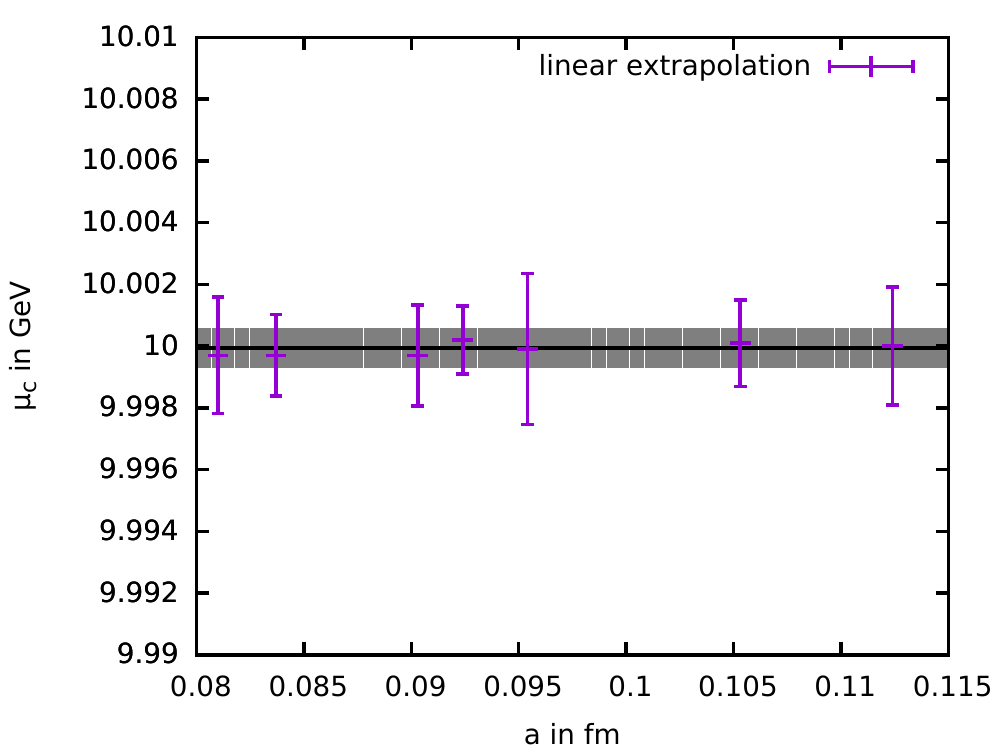}
 \caption{Critical chemical potentials $\mu_c$ for the diquark-density onset at $T=0$ from linear extrapolations for different lattice spacings between $a=0.0810$~fm and $0.1124$~fm corresponding to lattice couplings between $\beta = 2.5$ and $2.4$. \label{spacings}}

\end{minipage} \hspace*{0.1cm}
\begin{minipage}{0.49\textwidth}
	\vspace*{-0.4cm}
	\hskip -.1cm
 \includegraphics[width=\linewidth]{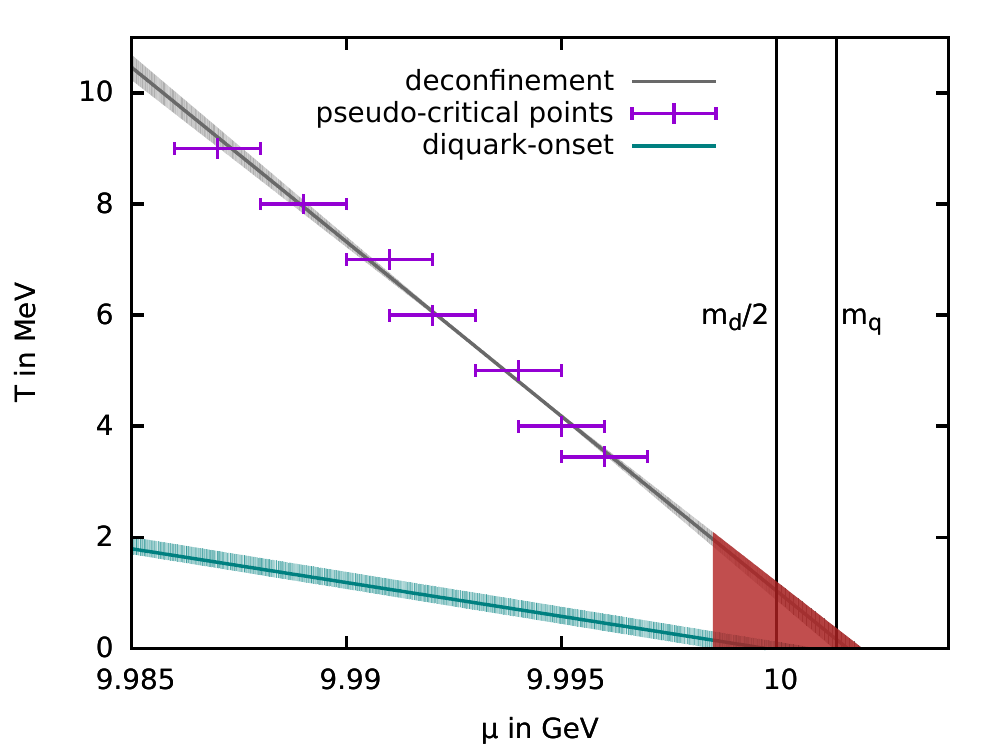}%
 \caption{Extrapolations of the deconfinement transition and the diquark-density onset (at $\beta = 2.5$), a possible superfluid diquark-condensation phase is marked by the shaded red triangle,  $m_d/2 = 10$~GeV and the quark mass $m_q = 10.0014$~GeV are indicated by vertical lines. \label{region}}
\end{minipage}
        \end{figure}
        
In order to test the scaling of this onset we have performed the same analysis also for 7 different lattice couplings $\beta $ between 2.4 and 2.5, corresponding to lattice spacings between $a= 0.1124$~fm and $0.0810$~fm with $\kappa$ values adjusted so that $m_d$ remains fixed at 20~GeV as before. Again, for each $\beta$ we extract the corresponding intersection points of the two exponential regimes in the quark density at the same 7 temperatures between 9~MeV and 3.454~MeV. The extrapolated $N_f=2$ results for the zero-temperature diquark-density onsets from these intersection points are collected in Fig.~\ref{spacings}. Within the errors, these extrapolated values for $\mu_c$  basically all agree. Assuming that $\mu_c$ is thus independent of the lattice spacing in this parameter regime we simply use their average as our final overall estimate of
\begin{equation}
  \mu_c = 9.9999(7)\; \mbox{GeV}
\end{equation}
from the data in Fig.~\ref{spacings} as indicated by the horizontal line with the gray error band. This overall estimate thus confirms that $\mu_c = m_d/2$ with rather high precision. Unfortunately, the region where one might find a superfluid diquark-condensation phase is currently still beyond reach within the convergence region of the hopping series. The temperatures from 3.5 MeV upwards are still too high for the 20 GeV diquarks which are bound by only about $2 m_q - m_d \approx 2.8$~MeV.

To give a rough estimate where a diquark superfluid might be found, we have also extrapolated the inflection point of the Polyakov loop as an indication of
deconfinement to $T=0$. Fig.~\ref{region} shows the phase diagram with the extrapolations around $\mu=m_d/2$.
As seen in the figure, the deconfinement crossover then hits the $T=0$ axis of the phase diagram just above $m_q = 10.0014$~GeV. Therefore, a small window for a potential superfluid diquark-condensation phase at sufficiently low temperatures remains. The region where this might occur is indicated by the shaded red triangle in Fig.~\ref{region}. It starts at the lower limit given by the extrapolation error of the diquark-density onset for the $\beta=2.5$ data used here. The deconfinement temperature at this lower limit is then of roughly the same order as the diquark-binding energy and hence a naive estimate of the transition temperature of a possible diquark-condensation phase.

\section{Effective Theory for G$_2$-QCD}
\begin{figure}[t]
\begin{minipage}{0.49\textwidth}
	\vspace*{-1.6cm}
	\hskip -.4cm
	\includegraphics[width=1.2\linewidth]{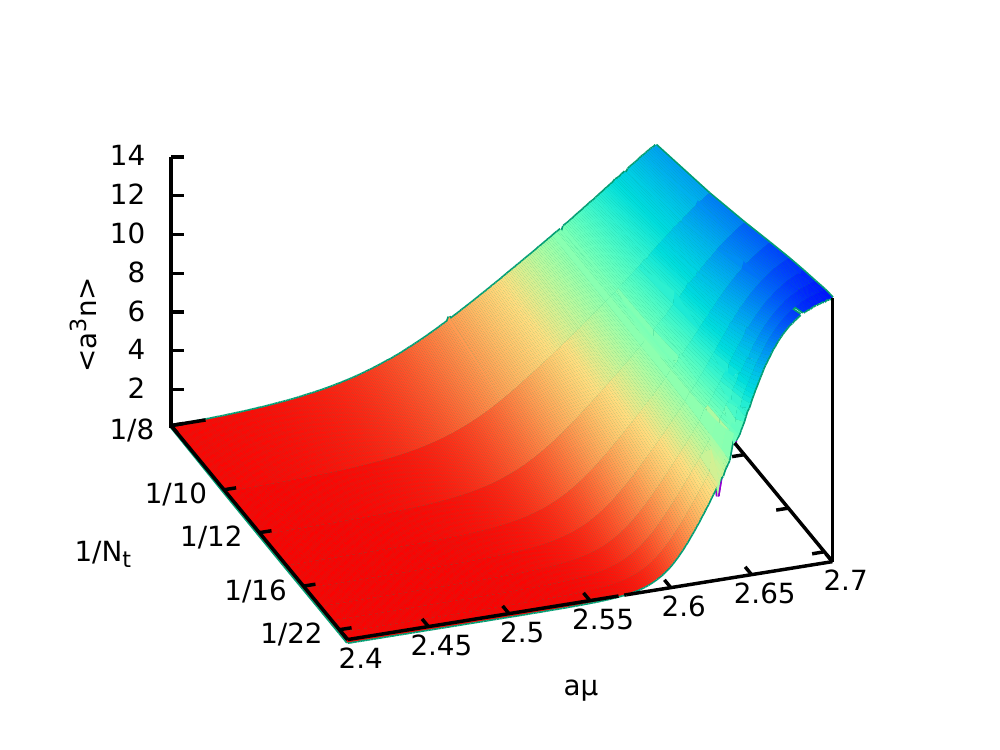}
 \caption{$G_2$-QCD quark density $a^3n$ in lattice units in the $\mu$-$1/N_t$ plane with simulation parameters $\beta/N_c=1.39$, $\kappa=0.0357$, $N_t=8,10,\dots,24$. \label{dens_g2}}

\end{minipage} \hspace*{0.1cm}
\begin{minipage}{0.49\textwidth}
	\vspace*{-0.6cm}
	\hskip -.2cm
 \includegraphics[width=1.1\linewidth]{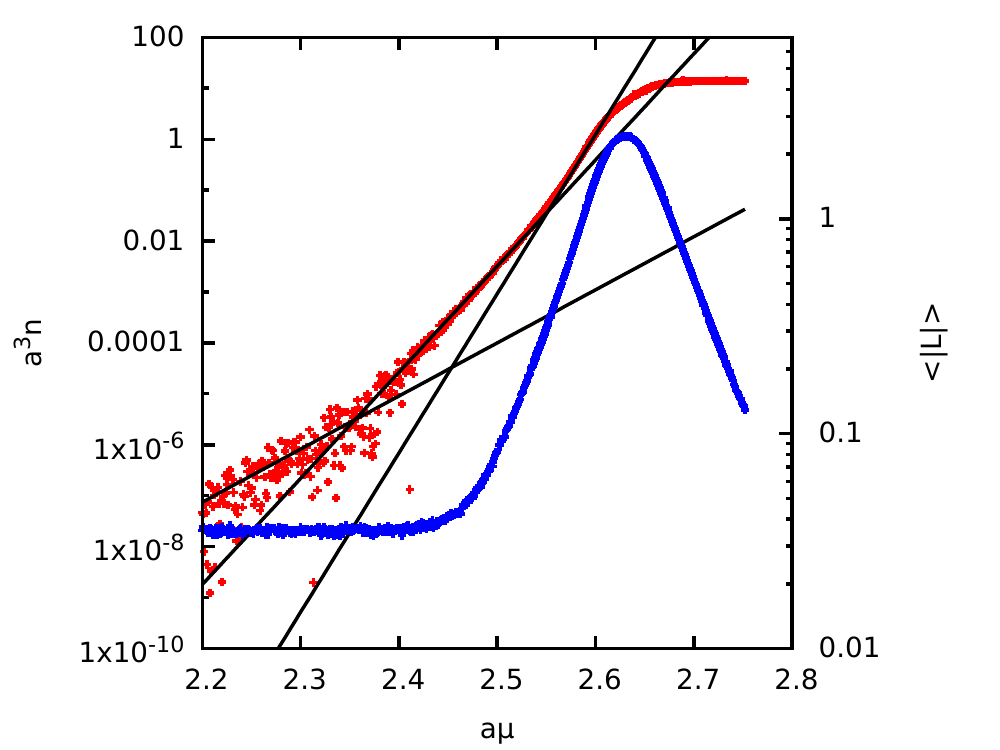}%
 \caption{Logarithmic plot of $G_2$-quark density $a^3n$ and Polyakov loop $\langle |L|\rangle $ over quark chemical potential for the parameters of Fig.~7 for $N_t=24$, three exponential regimes are indicated by solid lines. \label{log_g2}}
\end{minipage}
\vspace*{-.4cm}        

\end{figure}

We have also applied the effective theory to G$_2$-QCD with one flavor of Wilson quarks. Most of the steps to derive the effective action proceed completely analogous to two-color QCD. The only modifications to \eqref{eff_action_kappa4_Nf2} are the color and flavor factors, and the fact that we can drop the term in the last line of \eqref{eff_action_kappa4_Nf2}, since the corresponding diagrams in G$_2$-QCD only lead to an irrelevant constant. Figures \ref{dens_g2} and \ref{log_g2} show our preliminary results. Fig.~\ref{dens_g2} shows the quark density $a^3n$ in the $\mu$-$1/N_t$ plane of the phase diagram. We can see explicitly that the density increases with $\mu$ more rapidly as we lower the temperature. We again observe lattice saturation at the expected value $a^3n=2\cdot N_c=14$. Similar to our procedure for two-color QCD, Fig.~\ref{log_g2} shows a logarithmic plot of density and the Polyakov loop at $aT=1/24$. We can now identify three different exponential regimes, however. The heavy quark gas with imperfect confinement, a diquark gas $\propto \exp\{2\mu/T\}$, and a third exponential increase $\propto \exp\{3\mu/T\}$ corresponding to a finite density of $G_2$-nucleons.

\section{Conclusion and Outlook}
We have shown results for the cold and dense regions in the phase diagrams of two-color and $G_2$-QCD from effective Polyakov loop theories for heavy quarks. We observe regimes in which the densities behave as thermal gases of color singlet baryons with good evidence of a continuous transition to finite diquark density at zero temperature. The onset of diquark density in two-color QCD for $T \to 0$ extrapolates to $\mu_c = m_d/2$ with very high precision indicating that there is no binding energy as in nuclear matter in QCD. The effective theory for heavy quarks appears to reflect these essential differences between 2 and 3 colors \cite{Fromm:2012eb}. In the near future we will compare the results from the effective theory to full two-color and $G_2$-QCD simulations at finite density and assess the range of applicability and the quantitative accuracy of the effective theory in more detail. In $G_2$-QCD the additional question arises whether one can seperate the extrapolated zero-temperature onsets of diquark and nucleon density in the effective theory with smaller quark masses.

\end{document}